\documentstyle[12pt]{article}

\begin{document}
\title{Unitarity Restoration for the Product of Nonunitary Operators}

\author{Arlen Anderson\thanks{email: arley@physics.unc.edu}\\
Dept. Physics\\
University of North Carolina\\
CB\# 3255\\
Chapel Hill NC 27599-3255}

\date{Feb. 25, 1995}
\maketitle

\vspace{-9cm}

\hfill gr-qc/9503002

\vspace{9cm}

\begin{abstract}
A proof is given that the polar decomposition procedure for unitarity
restoration works for products of invertible nonunitary operators.  A
brief discussion follows that the unitarity restoration procedure, applied
to propagators in spacetimes containing closed timelike curves, is
analogous to the original introduction by Feynman of ghosts to restore
unitarity in non-abelian gauge theories.  (The substance of this
paper will be a note added in proof to the published version of
gr-qc/9405058, to appear in Phys Rev D.)
\end{abstract}

\newpage

In Ref.~\cite{And}, a polar decomposition procedure for restoring
unitarity to evolution by
invertible nonunitary operators was given (see also the discussion in
Ref.~\cite{FeW}).
This procedure is relevant to
understanding apparently nonunitary evolution in the presence of
closed timelike curves\cite{Bou,Har}.  A naive calculation suggests
that this procedure does not work for products of such operators\cite{Haw}.
Consider the composition of two invertible non-unitary operators
$X_2$ and $X_1$
\begin{equation}
X_{21}=X_2 X_1.
\end{equation}
Let $U_1=(X_1 X_1^\dagger)^{-1/2}X_1$, $U_2=(X_2
X_2^\dagger)^{-1/2}X_2$,  then it is easy to check that
\begin{equation}
U_{21}=(X_{21} X_{21}^\dagger)^{-1/2}X_{21}\ne U_2 U_1.
\end{equation}
It would seem that the unitarity restoration procedure has failed.
Fortunately, this argument rests on a couple of false assumptions,
principally involving where each operator is defined.

Let the initial Hilbert space be ${\cal H}_a$.  The key to the
unitarity restoration procedure in Ref.~\cite{And} is that $X_1$ is not
to be viewed as a
nonunitary operator from ${\cal H}_a$ to itself,
but as a unitary operator from ${\cal H}_a$ to some ${\cal H}_b$. This
means that $X_2$ should not be thought of as an operator acting on ${\cal
H}_a$ but as one acting on ${\cal H}_b$.  In particular, when making the
polar decomposition of $X_2$, one should have made it in ${\cal H}_b$,
using the appropriate adjoint $\dagger_b$.
But then $U_2$ would be a unitary operator in ${\cal H}_b$ and could not
be directly composed with $U_1$.   This shows why the above computation is
wrong.  The correct computation involves an additional subtlety.

If one does not want to change Hilbert spaces under evolution but instead
wants to remain in ${\cal H}_a$, this can be done.
There is a polar decomposition of $X_1$ in ${\cal H}_a$
\begin{equation}
X_1=R_1 U_1,
\end{equation}
where
\begin{equation}
\label{r1}
R_1=(X_1 X_1^{\dagger_a})^{1/2}=R_1^{\dagger_a}.
\end{equation}
Then $U_1=(X_1 X_1^{\dagger_a})^{-1/2}X_1$ is a unitary operator from
${\cal H}_a$ to itself. The unitarity restoration procedure characterized
by adjusting the final Hilbert space
so that $X_1$ is a unitary operator is equivalent to that of simply using
the unitary part of its polar decomposition $X_1$ in ${\cal H}_a$\cite{And}.

The measure density $\mu_b$ in ${\cal H}_b$
is related to $\mu_a$ in ${\cal H}_a$ by\cite{And}
\begin{equation}
\label{meas}
\mu_b=X_1^{-1\,\dagger_1} \mu_a X_1^{-1}= \mu_a (X_1
X_1^{\dagger_a})^{-1}= \mu_a R_1^{-2},
\end{equation}
where $\dagger_1$ means the adjoint in the trivial density $\mu=1$. By
virtue of this\cite{And}, $R_1$ can be understood as a map from ${\cal
H}_a$ to ${\cal H}_b$, and it is the operator which enables one to change
between the two descriptions of the unitarity restoration procedure. As an
operator connecting the two Hilbert spaces, $R_1$ can be used to pull back
operators $B$ which act on ${\cal H}_b$ to operators $\tilde B$ which act on
${\cal H}_a$
\begin{equation}
\tilde B=R_1^{-1}BR_1.
\end{equation}
As well, from Eq.~(\ref{meas}), $R_1^2$ is seen to transform adjoints
defined with respect to one
measure density to those defined with respect to the other, e.g.
\begin{equation}
\label{adjtr}
B^{\dagger_b} R_1^2= R_1^2 B^{\dagger_a}.
\end{equation}

The product $X_{21}=X_2 X_1=X_2 R_1 U_1$ can be read as unitary
evolution in ${\cal H}_a$ followed by the non-unitary operator $X_2 R_1$.
To make this non-unitary operator the pull-back of an operator on
${\cal H}_b$ to one on ${\cal H}_a$,
one decomposes $X_2=R_1^{-1} Y_2$ so that $X_2 R_1=R_1^{-1}
Y_2 R_1$.  The polar decomposition of $Y_2$ in ${\cal H}_b$ is
\begin{equation}
Y_2=R_2 U_2,
\end{equation}
where
\begin{equation}
\label{r2}
R_2=(Y_2 Y_2^{\dagger_b})^{1/2}
\end{equation}
and $U_2 U_2^{\dagger_b}=1$.  Then $R_1^{-1} Y_2 R_1$ contains the pull-back
$\tilde U_2=R_1^{-1} U_2 R_1$ to ${\cal H}_a$
of the unitary operator $U_2$ on ${\cal H}_b$.  Using Eq.~(\ref{adjtr})
applied to $U_2^{\dagger_b}$, it is
straightforward to verify that $\tilde U_2$ is unitary in ${\cal H}_a$.

The correct unitary composition law can now be derived using the
polar decomposition procedure, beginning from
\begin{eqnarray}
U_{21}&=&(X_{21} X_{21}^{\dagger_a})^{-1/2}X_{21} \\
&=& (R_1^{-1} Y_2 X_1 X_1^{\dagger_a} Y_2^{\dagger_a}
R_1^{-1\,\dagger_a})^{-1/2} R_1^{-1} Y_2 X_1. \nonumber
\end{eqnarray}
To simplify the square-root, one successively uses Eq.~(\ref{r1}),
Eq.~(\ref{adjtr}) applied to $Y_2^{\dagger_a}$, and Eq.~(\ref{r2}) to obtain
\begin{equation}
(R_1^{-1} R_2^2 R_1)^{-1/2} =R_1^{-1} R_2^{-1} R_1.
\end{equation}
Thus, one has
\begin{eqnarray}
U_{21}&=&(R_1^{-1} R_2^{-1} R_1) R_1^{-1} Y_2 X_1 \nonumber \\
&=& \tilde U_2 U_1.
\end{eqnarray}
This is the correct unitary composition law.  The unitarity restoration
procedure works fine on products of invertible nonunitary operators when
one is careful with the Hilbert spaces on which each operator acts.

In closing, I would like to offer a discussion of this unitarity
restoration procedure which was not given in Ref.~\cite{And}. The polar
decomposition version seems {\it ad hoc}, and to a large extent it is,
because one simply drops the troublesome non-unitary part of the evolution
operator. In the case of evolution in the presence of closed timelike
curves we are dealing with an unfamiliar mathematical problem. In
Ref.~\cite{Bou}, a natural path integral procedure was used to derive
propagators, yet while correct so far as they go, these calculations are
not necessarily complete. The unitarity restoration procedure gives the
outcome of a fairly minimal completion.

One recalls that in the early sixties Feynman\cite{Fey} introduced ghosts
to cancel an apparent non-unitarity in a perturbative treatment of
non-abelian gauge theories, and their success in this role was confirmed
through further calculations by DeWitt\cite{Dew}. Only later was an
understanding of these ghosts found through Faddeev-Popov
gauge-fixing\cite{FaP}. It may well be that some subtlety like ghosts is
involved in a careful path integral treatment of evolution in the presence
of closed timelike curves. For instance, one knows that ghosts generally
arise when enforcing constraints while the requirement of consistency in
the presence of closed timelike curves produces a timelike periodic
constraint on states. This constraint is not explicit in Ref.~\cite{Bou} but
is assumed to be taken care of automatically through the use of periodic
paths. That may not be sufficient.

As well, it is suggestive that the unitarity restoration procedure can be
understood in terms of a changing measure density under evolution.  Such a
change in measure density would modify the path integral to be computed
for the propagator.  Whether such a modification could be implemented
is not clear and requires study of specific examples.
Clearly, one should not rest with the unitarity restoration procedure
discussed here.  Further effort is necessary to find a more fundamental
justification for it.

Acknowledgement.  This work was supported in part by National
Science Foundation grant PHY 94-143207.

\end{document}